\documentclass{elsart}
\usepackage{epsfig}
\usepackage{colordvi}

\newcommand{\be}{\begin{equation}}
\newcommand{\ee}{\end{equation}}
\newcommand{\bea}{\begin{eqnarray}}
\newcommand{\eea}{\end{eqnarray}}
\newcommand{\bqa}{\begin{eqnarray}}
\newcommand{\eqa}{\end{eqnarray}}

\newcommand{\ep}{\varepsilon}

\def \litwo {{\rm{Li_2}}}
\def \litr {{\rm{Li_3}}}
\def \lifo {{\rm{Li_4}}}
\def \ep   {\epsilon}
\def \litt {{\rm{S_{2,2}}}}
\def \liot {{\rm{S_{1,2}}}}

\begin{document}

\begin{flushleft}
{\normalsize \tt
DESY 06-021
\\
WUE-ITP-2006-004
\\
SFB/CPP-06-17
\\ April 2006}
\end{flushleft}
\vspace*{1.cm}
\begin{frontmatter}
  \title {\bf The planar four-point master integrals \\
  for massive two-loop Bhabha scattering}
  \author{M. Czakon}
  \address{Institut f\"ur Theoretische Physik
    und Astrophysik, Universit\"at W\"urzburg, \\
    Am Hubland, D-97074 W\"urzburg, Germany, \\
    Department of Field Theory and Particle Physics,
    Institute of Physics, \\
    University of Silesia, Uniwersytecka 4, PL-40-007 Katowice,
    Poland}
  \author{J. Gluza}
    \address{Department of Field Theory and Particle Physics,
    Institute of Physics, \\
    University of Silesia, Uniwersytecka 4, PL-40-007 Katowice,
    Poland}
\author{T. Riemann}
\address{Deutsches Elektronen-Synchrotron, DESY, \\
   Platanenallee 6, 15738 Zeuthen, Germany}

\begin{abstract}
We present the values of the complete set of planar four-point Master
Integrals needed for massive Bhabha scattering in the limit of
fixed angle and high energy at the two-loop level. The integrals have
been calculated using direct expansions of Mellin-Barnes
representations, followed by a resummation of resulting harmonic
series.
\end{abstract}

\end{frontmatter}

\section{Introduction}

The Bhabha scattering process, {\it i.e.} the elastic scattering of an
electron and a positron is an invaluable tool in the luminosity
determination in various experiments. Devices like the International
Linear Collider require a theoretical prediction for the cross
section that contains, besides exponentiated factors coming from
the leading logarithmic terms, also fixed order contributions, and in
particular the two-loop QED corrections.

As far as large angle Bhabha scattering is concerned a prediction to
leading order in the small electron mass (we shall later refer generally to
any fermion species) has been derived in \cite{Penin}, although the
logarithmic terms have been known before \cite{Logs}. This prediction has
been obtained from the massless result \cite{Bern} by a careful matching
procedure. Since heavy use has been made of results from different
sources, it is definitely desirable to obtain the cross section by an
independent method. Even though much progress has been made on the way
to a result with an exact mass dependence \cite{Italians,NPB,PRD}, we are still
far from the goal. In the mean time, we have found it much easier
to evaluate the required diagrams in the same approximation as in
\cite{Penin}. The present paper presents our results for the set of planar
Master Integrals (MI). The main tool are expansions of Mellin-Barnes (MB)
integrals.

In the next section we present some considerations on the choice of
the MIs, we then describe the derivation and expansion of MB
integrals. We subsequently give our results and finish with
conclusions. An appendix contains our new MB representations for the
five-line integrals.

\section{Master Integrals}

As explained in the introduction, it is our purpose to
calculate the integrals in the limit of small fermion mass, which
corresponds to high-energy scattering at fixed angle. Since this is a
Sudakov approximation, we have to expect an overlap of collinear and
on-shell infrared singularities leading to logarithms of the fermion
mass reaching power four at the two-loop level. Even though the final
result for the cross section will only be given at leading order, {\it
i.e.} neglecting terms suppressed by $m^2$, it is not immediately
clear how many terms of the expansion of the individual Master
Integrals are needed. Indeed, it is possible that some MIs appear
with such coefficients that terms of relative order $1/m^2$ would
result. These would, of course, have to cancel at the end but they
would also make a higher order expansion in $m^2$ necessary. One way
to be sure that the choice of masters avoids this situation is simply
to calculate the amplitude and find a set of MIs, whose coefficients
would all be of relative order $O(1)$ in the mass expansion. An
alternative strategy, which can be applied without knowing the full
amplitude is to require that the integrals have their mass dimension
given by the kinematic invariants alone. To be specific, we shall
require that for any integral $I$, the following relation holds:
\be
\label{requirement}
I = (m^2)^{-2\epsilon} s^n F\left(\ln\left( -\frac{m^2}{s}\right),
\frac{t}{s}\right)+O\left(m^2\right),\;\;\;\; n =
\frac{1}{2}\;\mbox{dim}\; I+2\epsilon,
\ee
where $s$ and $t$ are the usual Mandelstam kinematic invariants,
$\epsilon$ is defined through the dimension of space-time in dimensional
regularization, $d = 4-2\epsilon$, and $F$ is some function that we
wish to determine. The factor $(m^2)^{-2\epsilon}$ could have been
replaced by $(-s)^{-2\epsilon}$ through a reshuffling of logarithms,
but we shall keep it as it is, because at intermediate stages of our
calculation the mass has been set to unity for simplification.

Whether Eq.~(\ref{requirement}) can be satisfied for a given integral
or not depends entirely on the strength of the IR
singularities. Indeed, usual Feynman graphs have only logarithmic
branching points, but putting a dot on some line may result in
stronger divergences. This is easily seen in a graph that would have
just one massive line regulating the collinear divergence. Putting a
dot on it, would be equivalent to a derivative, which transforms a
logarithm $\ln^n(-m^2/s)$ into $n \ln^{n-1}(-m^2/s)/m^2$, which is
precisely what we want to avoid. Of course, choosing MIs with property
Eq.~(\ref{requirement}) is not yet a proof that there will be no
coefficients with $1/m^2$, but we will be satisfied with the
assumption, based on experience, that this is indeed the case.

The consequence of the above considerations is that we have to slightly
modify the set of integrals with respect to the original one presented
in \cite{PRD}. In particular, most of the dotted masters have to be
replaced by those with irreducible numerators. The new, equivalent
set, is given in Section~\ref{results}. We have, of course, checked that
the listed integrals are indeed algebraically independent. In case the
reader would like to move back to the old set, we provide the
transition formulae in \cite{WWW}.

Having chosen the set of integrals to calculate, there is another
question to be considered,  namely whether one needs to derive the
expansions of the integrals both in the $s$ and in the $t$ channel, as
they both occur in the Bhabha scattering amplitude. Fortunately, the
limit under study is $s,t \gg m^2$, which means that the results are
valid in both cases. For consistency of notation, one just has to
change
\be
\ln\left(-\frac{m^2}{s}\right) \longrightarrow
\ln\left(-\frac{m^2}{t}\right) = \ln\left(-\frac{m^2}{s}\right)
-\ln\left(\frac{t}{s}\right),
\ee
and
\be
\frac{t}{s} \longrightarrow \frac{s}{t} = 1/\left(\frac{t}{s}\right).
\ee
The above transformations imply some algebra in transforming the
arguments of the polylogarithms to a unique form. In order to spare
the reader the effort, we provide the expansions in both channels in
\cite{WWW}.

\section{Expansion of Mellin-Barnes representations}

Mellin-Barnes representations for massive box integrals with the
highest number of lines (seven) at the two-loop level  have been
derived in \cite{Smirnov1,Smirnov2}. Since general powers of the
propagators have been kept, it should, in principle, be possible to
obtain representations for any box integral with a smaller number of
lines. The occurrence of factors $1/\Gamma(a)$, where $a$ is a
propagator power, seems to make some of the MB integrals trivially vanish.
This is actually not the case, since they turn out to be singular in
$a$ and an analytic continuation in the relevant propagator power
leads to a non-zero result with a smaller dimension of the integral.
This procedure is not always optimal and it turns out to be better to
derive MB integrals of smaller dimensionality directly for the
considered graph. This is particularly true for the 5-line integrals.

There is another case, where direct derivation of representations is
necessary. We have seen in the previous section that we need master
integrals with irreducible numerators. In such cases it is impossible
to obtain the MB integrals directly from the results for the
7-liners. In any case, we derived our own representations by the
standard technique of loop-wise integration, with Feynman parameter
representations for tensors as given for example in \cite{PRD}. The new
integrals can be found in the Appendix.

Having a MB integral at hand, one has to perform an analytic
continuation in $\epsilon$ from a range where the integral is regular
to the vicinity of the origin, uncovering the singularity structure on
the way. We perform this operation with the {\tt MATHEMATICA} package
{\tt MB} \cite{MB}. As a result, we obtain multidimensional integrals,
which still have a nontrivial dependence on both Mandelstam
variables. At this stage, the integral can be cast into the following
form \be I = (m^2)^{-2\epsilon}\int_{-i \infty}^{i \infty} d z \;
\left(-\frac{m^2}{s}\right)^z f\left(\frac{t}{s},z\right), \ee where
the $f$ function contains, amongst others, a product of $\Gamma$, or
possibly $\psi$ functions, which have poles in $z$. Let us stress once
more that, bar trivial cases, the $f$ function is given by a
multidimensional integral. Since it is difficult to directly take
residues in this form, we change the order of integration and close
the $z$ contour to the right. This procedure is subsequently applied
recursively, until no further poles at the required order of expansion
occur. Since the {\tt MB} package can perform numerical integrations,
we could check some of the integrals beyond the leading order against
the sector decomposition method with relatively large fermion
masses. Moreover, we could confirm explicitly that the expansion is
given by powers of $m^2$ and not of $m$, even though odd powers of the
latter are found in individual terms of the series.

Interestingly, the nontrivial dependence on the $t/s$ ratio occurred at
the end only in one-dimensional integrals. This is in strong contrast
to the massless calculation in \cite{SmirnovMassless}, for example. The
one-dimensional integrals were all of the harmonic type and could be
done with {\tt XSUMMER} \cite{Moch}. The remaining, constant, integrals
contained $\Gamma$ functions of doubled argument, $\Gamma(a\pm2 z)$,
which is a trace of the original mass. These have been dealt with
high precision numerical integration followed by the use of
the PSLQ algorithm \cite{PSLQ}. The transcendental constants that could
occur in the final result had to be Riemann $\zeta$ numbers up to weight
four, because of the correspondence between the expanded massive and
the purely massless result.

\section{Results}

\label{results}

All our integrals are defined with the integration measure
\be
\left(\frac{e^{\epsilon \gamma_E}}{i \pi^{2-\epsilon}}\right)^2 \int\int
  d^d k_1\;d^d k_2.
\ee
In the presentation of the results, we will use the following notation
\be
x=\frac{t}{s},\;\;\;\; L = \ln\left(-\frac{m^2}{s}\right),
\ee
which guarantees that our results are explicitly real in the
Euclidean domain, $s,t < 0$; and
\be
{\rm{S}}_{n,p} (z) = \frac{(-1)^{n+p-1}}{(n-1)!p!} ~ \int_{0}^{1}
\frac{dt}{t} \ln^{n-1}(t) \ln^{p}(1-zt),\;\;\;\;
{\rm{Li}}_n (z) = {\rm{S}}_{n-1,1} (z),
\ee
where $S_{n,p}(z)$ is the Nielsen polylogarithm. We will, furthermore,
always neglect the $(m^2)^{-2\epsilon}$ factor in front of the
integrals, and keep the $u$ Mandelstam variable in the prefactors,
with the understanding that
\be
s+t+u = 0,
\ee
which is correct in the limit under consideration. In fact, we believe
that the difficulties encountered in the exact evaluation of the
5-line masters in our previous works are precisely connected to the
occurrence of the $u$ variable in the prefactors, which breaks the
simplifications brought by the use of the conformally mapped variables
\be
X=\frac{\sqrt{1-4 m^2/s}-1}{\sqrt{1-4 m^2/s}+1},\;\;\;\;
Y=\frac{\sqrt{1-4 m^2/t}-1}{\sqrt{1-4 m^2/t}+1} .
\ee

\subsection{7-line integrals}

\begin{figure}
  \begin{center}
    \epsfig{file=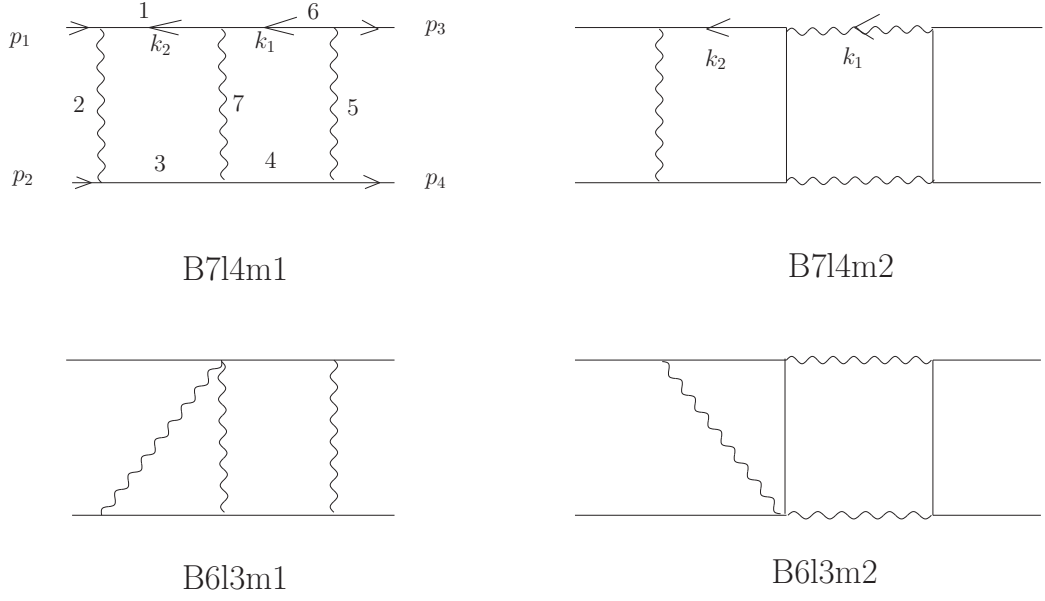,width=14cm}
  \end{center}
  \caption{\label{7lin}
The planar 6- and 7-line topologies. The momentum distribution is
common to all integrals.
}
\end{figure}

The two 7-line topologies are given in Fig.~\ref{7lin}. Here and in
the following we will give the irreducible numerator in parentheses
after the integral name. For the 7- and 6-liners we introduce the
shorthand
\be
N=(k_2+p_4)^2,
\ee
for the irreducible numerator as in \cite{Smirnov1,Smirnov2}. For the
calculation we used the MB representation from the mentioned papers.

\begin{eqnarray}
{\tt B7l4m1} =
       & + & \frac{1}{\epsilon^2}\; \frac{2 {\rm L}^2}{s^2t}
\nonumber \\
        &-& \frac{1}{\epsilon}\; \frac{1}{3 s^2t}
 \left[-10\;{\rm L}^3 + 12\;{\rm L}^2\;\ln(x) + 6\;{\rm L}\;\zeta_2 + 6\;\zeta_3  \right] \nonumber \\
        &-& \frac{1}{6s^2t} \biggl\{
-12\;{\rm L}^4 + 28\;{\rm L}^3\;\ln(x) - 4\;{\rm L}^2\;(-30\;\zeta_2 +
3\;\ln^2(x)) \nonumber \\
&-& 4\;{\rm L}\;(9\;\zeta_3 + 30\;\zeta_2\;\ln(x) + \ln^3(x) -
18\;\zeta_2\;\ln(1 + x) \nonumber \\
&-&
3\;\ln^2(x)\;\ln(1 + x) - 6\;\ln(x)\;{\litwo}( -x) + 6\;{\litr}( -x))
\nonumber \\
&-&15 \zeta_4-24 \zeta_3 \ln(x) \biggr\}
\label{B7l4m1ms}
\end{eqnarray}
\vspace*{-.6cm}
\begin{eqnarray}
{\tt B7l4m1(N)} =
& + & \frac{1}{\epsilon^2}\; \biggl[
 \frac{3}{2s^2} {\rm L}^2 \biggr] \nonumber \\
 & + & \frac{1}{\epsilon}\; \frac{1}{s^2} \biggl[
13/6\;{\rm L}^3 - 2\;{\rm L}^2\;\ln(x)- 2\;{\rm L}\;\zeta_2 - \zeta_3  \biggr]
\nonumber \\
 &+& \frac{1}{24s^2} \biggl\{
29\;{\rm L}^4 - 48\;{\rm L}^3\;\ln(x) - 408\;{\rm L}^2\;\zeta_2 \nonumber \\
&+& {\rm L}\;[168\;\zeta_3 + 384\;\zeta_2\;\ln(x) + 16\;\ln^3(x) -
288\;\zeta_2\;\ln(1 + x) \nonumber \\
&-& 48\;\ln^2(x)\;\ln(1 + x) - 96\;\ln(x)\;{\litwo}( -x)
+ 96\;{\litr}( -x)] + 144\;\zeta_4
\biggr\}
\label{B7l4m1Nms}
\end{eqnarray}

\begin{eqnarray}
{\tt B7l4m2}  =
 & -  & \frac{1}{\epsilon^2}\;
  \frac{1}{s^2t} \biggl[-{\rm L}^2 + {\rm L} \;\ln(x) \biggr]
\nonumber \\
 &  - & \frac{1}{\epsilon}\;
  \frac{1}{s^2t} \biggl[
-2\;{\rm L}^3 + 3\;{\rm L}^2\;\ln(x) - {\rm L}\;\ln(x)^2 - \zeta_2\;\ln(x)
\biggr] \nonumber \\
 &-& \frac{1}{6 s^2t}  \biggl\{
-10\;{\rm L}^4 + 26\;{\rm L}^3\;\ln(x) - 2\;{\rm L}^2\;(-21\;\zeta_2 +
9\;\ln^2(x)) \nonumber \\
&-& 2\;{\rm L}\;[30\;\zeta_3 + 21\;\zeta_2\;\ln(x) - \ln^3(x) -
18\;\zeta_2\;\ln(1 + x) \nonumber \\
&-&
3\;\ln(x)^2\;\ln(1 + x) - 6\;\ln(x)\;{\litwo}( -x) + 6\;{\litr}( -x)]
\nonumber \\
&+&  279 \;\zeta_4  + 18 \;\zeta_3 \;\ln(x) + 6 \;\zeta_2 \;\ln^2(x)
 \biggr\}
\label{B7l4m2ms}
\end{eqnarray}

\begin{eqnarray}
{\tt B7l4m2(N)}  = & +  & \frac{1}{\epsilon^2}\;
  \frac{1}{s^2} \biggl[ {\rm L}^2 - {\rm L} \;\ln(x) \biggr] \nonumber \\
&+& \frac{1}{\epsilon}\;
  \frac{1}{s^2} \biggl[ 2 \;{\rm L}^3 - 3 \;{\rm L}^2 \;\ln(x)+ {\rm L} \;\ln^2(x)+ \zeta_2 \;\ln(x)
\biggr] \nonumber \\
 &+& \frac{1}{24s^2} \biggl\{
25\;{\rm L}^4 - 48\;{\rm L}^3\;\ln(x) - 456\;{\rm L}^2\;\zeta_2  + {\rm L}\;[192\;\zeta_3 + 696\;\zeta_2\;\ln(x)
\nonumber \\
&+& 32\;\ln^3(x) - 288\;\zeta_2\;\ln(1 + x) - 48\;\ln^2(x)\;\ln(1 + x)
\nonumber \\
&-& 96\;\ln(x)\;{\litwo}( -x) + 96\;{\litr}( -x)] \nonumber \\
&-&804\;\zeta_4 - 168\;\zeta_3\;\ln(x) - 216\;\zeta_2\;\ln^2(x) -
8\;\ln^4(x) \nonumber \\
&+& 96\;\zeta_2\;\ln(x)\;\ln(1 + x) + 16\;\ln^3(x)\;\ln(1 + x) -
192\;\zeta_2\;{\litwo}( -x)
\nonumber \\
&+& 96\;\ln(x)\;{\litr}( -x) - 192\;{\lifo}( -x)
 \biggr\}
\label{B7l4m2N1ms}
\end{eqnarray}

\begin{eqnarray}
{\tt B7l4m2(N^2)}  =
& -  & \frac{1}{\epsilon^2}\;  \;\frac{x}{s} \biggl[ -{\rm L}^2 + {\rm L} \;\ln(x) \biggr]
\nonumber \\
&-& \frac{1}{\epsilon}\;\frac{1}{s} \; \biggl[
-2\;{\rm L}^3\;x
+\;{\rm L}^2\;[3\;x\;\ln(x)-1/2]
-{\rm L}\;x\;\ln^2(x) -4\;\zeta_2-x\;\zeta_2\ln(x) \biggr] \nonumber \\
&-& \frac{1}{120s} \biggr\{
-125\;{\rm L}^4\;x - 5\;{\rm L}^3\;[16 + 56\;x - 48\;x\;\ln(x)] \nonumber \\
&-& 5\;{\rm L}^2\;[24 - 456\;x\;\zeta_2 - 144\;x\;\ln(x)] \nonumber \\
&-&
 5\;{\rm L}\;[120\;\zeta_2 + 384\;x\;\zeta_2 + 192\;x\;\zeta_3 + 696\;x\;\zeta_2\;\ln(x) + 96\;x\;\ln^2(x)
\nonumber \\
&+& 32\;x\;\ln(x)^3 - 288\;x\;\zeta_2\;\ln(1 + x) -
48\;x\;\ln^2(x)\;\ln(1 + x)
\nonumber \\
&-& 96\;x\;\ln(x)\;{\litwo}( -x) + 96\;x\;{\litr}( -x)]
-960\;\zeta_2 - 1920\;x\;\zeta_2 - 840\;\zeta_3 \nonumber \\
&+& 480\;x\;\zeta_3 + 4020\;x\;\zeta_4
+ 1440\;x\;\zeta_2\;\ln(x) + 840\;x\;\zeta_3\;\ln(x)\nonumber \\
& +& 1080\;x\;\zeta_2\;\ln^2(x) + 80\;x\;\ln^3(x)
+ 40\;x\;\ln^4(x) - 1440\;x\;\zeta_2\;\ln(1 + x) \nonumber \\
&-&
 480\;x\;\zeta_2\;\ln(x)\;\ln(1 + x) - 240\;x\;\ln^2(x)\;\ln(1 + x)
\nonumber \\
&-& 80\;x\;\ln(x)^3\;\ln(1 + x) + 960\;x\;\zeta_2\;{\litwo}( -x)
- 480\;x\;\ln(x)\;{\litwo}( -x) \nonumber \\
&+& 480\;x\;{\litr}( -x) -
 480\;x\;\ln(x)\;{\litr}( -x) + 960\;x\;{\lifo}( -x)
\biggr\}
\end{eqnarray}

\begin{eqnarray}
{\tt B7l4m2(N^3)} = & + & \frac{1}{\epsilon^2}\;
\;x^2 \biggl[ {\rm L}^2 - {\rm L}\;\ln(x) \biggr]  \nonumber \\
&+& \frac{1}{\epsilon} \; \biggl[
2\;{\rm L}^3\;x^2+ {\rm L}^2\;(-1/4 + x/2 - 3\;x^2\;\ln(x))
+ {\rm L}\;(1 - x + x^2\;\ln^2(x))
\nonumber \\
&-& 2\;\zeta_2 + 4\;x\;\zeta_2 + x^2\;\zeta_2\;\ln(x) \biggr]
\nonumber \\
&+&\frac{1}{24} \biggl\{
25\;{\rm L}^4\;x^2 + {\rm L}^3\;(-8 + 16\;x + 84\;x^2 - 48\;x^2\;\ln(x))
\nonumber \\
&+& {\rm L}^2\;(24 - 72\;x^2 - 456\;x^2\;\zeta_2 - 216\;x^2\;\ln(x))
\nonumber \\
&+&
 {\rm L}\;(120 - 120\;x - 60\;\zeta_2 + 120\;x\;\zeta_2 + 576\;x^2\;\zeta_2
+ 192\;x^2\;\zeta_3 \nonumber \\
&+& 96\;x^2\;\ln(x) + 696\;x^2\;\zeta_2\;\ln(x) + 144\;x^2\;\ln^2(x)
+ 32\;x^2\;\ln^3(x)
\nonumber \\
&-& 288\;x^2\;\zeta_2\;\ln(1 + x) -
   48\;x^2\;\ln(x)^2\;\ln(1 + x) \nonumber \\
&-& 96\;x^2\;\ln(x)\;{\litwo}( -x)
+ 96\;x^2\;{\litr}( -x)) \biggr\} \nonumber \\
&+&\biggl\{
(3\;((-6 + 22\;x + 60\;x^2)\;\zeta_2 + (-7 + 2\;(7 - 6\;x)\;x)\;\zeta_3
- 67\;x^2\;\zeta_4)
\nonumber \\
&+& 2\;x^2\;(-(\ln(x)\;(54\;\zeta_2 + 21\;\zeta_3 + \ln(x)\;(27\;\zeta_2
+ \ln(x)\;(3 + \ln(x)))))
\nonumber \\
&+&
  (9 + 2\;\ln(x))\;(6\;\zeta_2 + \ln^2(x))\;\ln(1 + x)))/6 \nonumber \\
&-& 2\;x^2\;((4\;\zeta_2 - 3\;\ln(x))\;{\litwo}( -x) \nonumber \\
 &+& (3 - 2\;\ln(x))\;{\litr}( -x)
+ 4\;{\lifo}( -x))
\biggr\}
\label{B7l4m2N3ms}
\end{eqnarray}

\subsection{6-line integrals}

The 6-line integrals are obtained from the 7-line topologies of
Fig.~\ref{7lin}, by removing line 1 for {\tt B6l3m1} and line 3 for
{\tt B6l3m2}, and keeping the same momentum distribution. This follows
from the fact that we use the same original MB representation from
\cite{Smirnov1,Smirnov2}.

\begin{eqnarray}
{\tt B6l3m1} =
& +  & \frac{1}{\epsilon}\; \frac{1}{2 st} \biggl[
{\rm L}^3-2 {\rm L}^2 \ln(x)+{\rm L} [8 \zeta_2 + \ln^2(x) ]
\biggr] \nonumber \\
&+&   \frac{1}{24st} \biggl\{
7 \;{\rm L}^4 - 32 \;{\rm L}^3 \;\ln(x) + {\rm L}^2 \;(24 \;\zeta_2 + 42 \;\ln^2(x)) \nonumber \\
&+& {\rm L} \;(24 \;\zeta_3 - 72 \;\zeta_2 \;\ln(x) - 16 \;\ln^3(x) + 72 \;\zeta_2 \;\ln(1 + x) \nonumber \\
&+& 12 \;\ln^2(x) \;\ln(1 + x) + 24 \;\ln(x) \;{\litwo}( -x) - 24 \;{\litr}( -x))
\biggr\}
\nonumber \\
&+& \frac{1}{24st} \biggl\{
-672 \;\zeta_4 - 48 \;\zeta_2 \;\ln^2(x) - \ln^4(x) - 12 \;(6 \;\zeta_2 + \ln^2(x)) \;{\litwo}( -x)
\nonumber \\
&+&
48 \;\ln(x) \;{\litr}( -x) - 72 \;{\lifo}( -x)
\biggr\}
\end{eqnarray}

\begin{eqnarray}
{\tt B6l3m1(N)}  =
& -  & \frac{1}{\epsilon}\; \frac{1}{2s} \biggl[
-{\rm L}^2-4 {\rm L}+2 \zeta_2 \biggr] \nonumber \\
&-& \frac{1}{6s} \biggl\{
{\rm L}^3  + {\rm L}^2 \;(-6 - 12 \;\ln(x))
\nonumber \\
&+& {\rm L} \;(-24 + 24 \;\zeta_2 + 6 \;\ln^2(x))
+ 12 \;\zeta_2 + 12 \;\zeta_3
\biggr\}
\end{eqnarray}

\begin{eqnarray}
{\tt B6l3m2}  =
&+&\frac{1}{24 st} \biggl\{
7 \;{\rm L}^4 - 28 \;{\rm L}^3 \;\ln(x) + 6 \;{\rm L}^2 \;(26 \;\zeta_2 + 7 \;\ln^2(x)) \nonumber \\
&+& {\rm L} \;(-312 \;\zeta_2 \;\ln(x) - 28 \;\ln^3(x) + 12 \;(6 \;\zeta_2 + \ln^2(x)) \;\ln(1 + x)
\nonumber \\
&+& 24 \;\ln(x) \;{\litwo}( -x) - 24 \;{\litr}( -x)) \nonumber \\
&+&
510 \;\zeta_4 + 156 \;\zeta_2 \;\ln^2(x) + 7 \;\ln^4(x) - 8 \;\ln(x) \;(6 \;\zeta_2 + \ln^2(x)) \;\ln(1 + x)
\nonumber \\
&+& 4 \;(6 \;\zeta_2 - 3 \;\ln^2(x)) \;{\litwo}( -x) + 24 \;{\lifo}( -x)
\biggr\}
\end{eqnarray}

\begin{eqnarray}
{\tt B6l3m2(N)}  =
 & -  & \frac{1}{\epsilon}\; \frac{1}{2s} \biggl[
-{\rm L}^2-8 \zeta_2 \biggr] \nonumber \\
 &-& \frac{1}{6s} \biggl\{
3 \;{\rm L}^3 + 3 \;{\rm L}^2 \;(-2 - 6 \;\ln(x)) + 3 \;{\rm L} \;(18 \;\zeta_2 + 6 \;\ln^2(x)) \biggr\} \nonumber \\
 &-& \frac{1}{s} \biggl\{
-9 \;\zeta_3 - 10 \;\zeta_2 \;\ln(x) - \ln^3(x) + 6 \;\zeta_2 \;\ln(1 + x) + \ln^2(x) \;\ln(1 + x)
\nonumber \\
&+& 2 \;\ln(x) \;{\litwo}( -x) - 2 \;{\litr}( -x) \biggr\}
\end{eqnarray}

\subsection{5-line integrals}

The topologies for the 5-liners with numerators are given in Fig.~\ref{5lin}. For the
evaluation of the integrals we used our own MB representations given
in the Appendix. The position of the dots on the dotted integrals
(having a ``d'' in their name) has been defined in \cite{PRD}. These
integrals belong to our original MI set.

\begin{figure}
  \begin{center}
    \epsfig{file=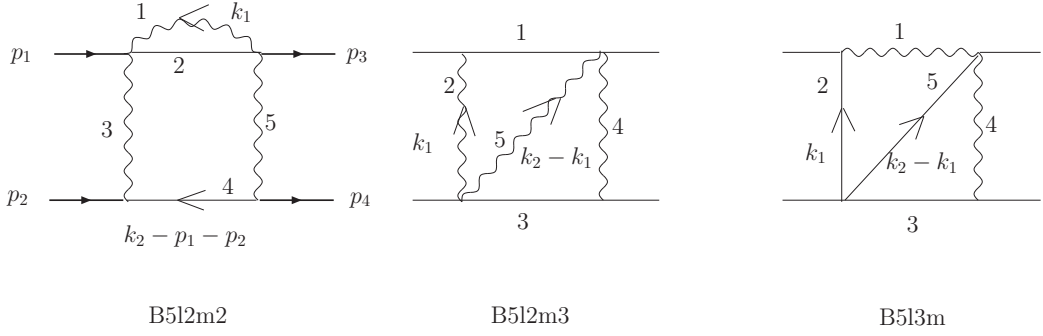,width=14cm}
  \end{center}
  \caption{\label{5lin}
The 5-line topologies. The momentum distribution has been chosen to
make the derivation of MB representations easier.
}
\end{figure}

\begin{eqnarray}
{\tt B5l2m1}  =
&+& \frac{1}{\epsilon^2} \frac{1}{s}\;[{\rm L}]
\nonumber \\
&-& \frac{1}{\epsilon} \; \frac{1}{s} \;\biggl[
-3/2 \;{\rm L}^2 + \;{\rm L}\;(-2 + \ln(x))+\;\zeta_2 \biggr]
 \nonumber \\
&-& \frac{1}{6s} \biggl\{
-7\;{\rm L}^3 + {\rm L}^2\;(-18 + 9\;\ln(x)) + {\rm L}\;(-24 + 24\;\zeta_2
+ 12\;\ln(x) - 3\;\ln^2(x))
\nonumber \\
&+&
12\;\zeta_2 + 12\;\zeta_3 - 24\;\zeta_2\;\ln(x) - \ln^3(x) +
18\;\zeta_2\;\ln(1 + x)
\nonumber \\
&+&  3\;\ln^2(x)\;\ln(1 + x) + 6\;\ln(x)\;{\litwo}( -x) - 6\;{\litr}( -x)
\biggr\}
\end{eqnarray}


\begin{eqnarray}
{\tt B5l2m2} =
& - & \frac{1}{\epsilon}\;
\frac{1}{6t} \biggl[
-3\;{\rm L}^2+6\;{\rm L} \;\ln(x) - 3 \;\ln^2(x)-24\;\zeta_2 \biggr]
\nonumber \\
&-& \frac{1}{6t} \biggl\{
-4\;{\rm L}^3 - {\rm L}^2\;(6 - 9\;\ln(x)) - {\rm L}\;(6\;\zeta_2 - 12\;\ln(x) + 6\;\ln^2(x)) \nonumber\\
&-&48\;\zeta_2 - 54\;\zeta_3 - 18\;\zeta_2\;\ln(x) - 6\;\ln^2(x)
+ \ln^3(x) \nonumber \\
&+& 18\;\zeta_2\;\ln(1 + x) + 3\;\ln^2(x)\;\ln(1 + x) \nonumber \\
&+& 6\;\ln(x)\;{\litwo}( -x) - 6\;{\litr}( -x)
\biggr\}
\label{B5l2m2ms}
\end{eqnarray}

\begin{eqnarray}
{\tt B5l2m2(k_2\cdot p_3)}\; &=& \; +  \frac{1}{\epsilon}\;
\frac{s}{12t} \biggl[
3\;{\rm L}^2 + {\rm L}\;(12 + 6\;x - 6\;\ln(x)) \nonumber \\
&+& 24\;\zeta_2 - 12\;\ln(x) - 6\;x\;\ln(x) + 3\;\ln^2(x)
\biggr]
\nonumber \\
&&+
\frac{s}{12t} \biggl\{
4\;{\rm L}^3 + {\rm L}^2\;(3\;(8 + 3\;x) - 9\;\ln(x)) \nonumber \\
&+& {\rm L}\;[6\;(10 + 5\;x + \zeta_2) - 12\;(3 + x)\;\ln(x)
+ 6\;\ln^2(x)] \nonumber \\
&+&12\;\zeta_2 + 54\;\zeta_3 - 60\;\ln(x) - 30\;x\;\ln(x) +
18\;\zeta_2\;\ln(x) + 12\;\ln^2(x)
\nonumber \\
&+& 6\;x\;\ln^2(x) - \ln^3(x) - 18\;\zeta_2\;\ln(1 + x) -
3\;\ln^2(x)\;\ln(1 + x)
\nonumber \\
&-& 6\;\ln(x)\;{\litwo}( -x) + 6\;{\litr}( -x)
 \biggr\}
\label{B5l2m2Nms}
\end{eqnarray}

\begin{eqnarray}
{\tt B5l2m2d2} = & + &
\frac{1}{6st} \biggl\{
-12\;{\rm L}^3  + 9\;{\rm L}^2\;\ln(x)- 36\;{\rm L}\;\zeta_2 \nonumber \\
&-&12\;\zeta_3 - 12\;\zeta_2\;\ln(x) - \ln^3(x) +
18\;\zeta_2\;\ln(1 + x) \nonumber \\
&+& 3\;\ln^2(x)\;\ln(1 + x) + 6\;\ln(x)\;{\litwo}( -x) - 6\;{\litr}( -x)
\biggr\}
\label{B5l2m2d2ms}
\end{eqnarray}

\begin{eqnarray}
{\tt B5l2m3} = & + &
\frac{1}{12 u} \biggl\{
-6\;{\rm L}^2\;(6\;\zeta_2 + \ln^2(x)) \nonumber \\
&-& 6\;{\rm L}\;(-4\;\zeta_3 + 4\;\zeta_2\;\ln(x) - 12\;\zeta_2\;\ln(1 + x) - 2\;\ln^2(x)\;\ln(1 + x)
\nonumber \\
&-& 4\;\ln(x)\;{\litwo}( -x) + 4\;{\litr}( -x)) \nonumber \\
&+&312\;\zeta_4 + 72\;\zeta_3\;\ln(x) + 36\;\zeta_2\;\ln^2(x) + \ln^4(x)
- 24\;\zeta_3\;\ln(1 + x)
\nonumber \\
&+& 24\;\zeta_2\;\ln(x)\;\ln(1 + x) - 36\;\zeta_2\;\ln^2(1 + x)
- 6\;\ln^2(x)\;\ln^2(1 + x)
\nonumber \\
&-& 24\;\ln(x)\;  \liot (-x) + 
 12\;(8\;\zeta_2 + \ln^2(x) - 2\;\ln(x)\;\ln(1 + x))
\;{\litwo}( -x) \nonumber \\
&-& 48\;\ln(x)\;{\litr}( -x) + 24\;\ln(1 + x)\;{\litr}( -x)
\nonumber \\
&+& 72\;{\lifo}( -x) + 24\;{\litt}( -x)
\biggr\}
\label{B5l2m3ms}
\end{eqnarray}

\begin{eqnarray}
{\tt B5l2m3d2} = & - &  \frac{1}{\epsilon^2}\;
 \frac{1}{st}  {\rm L}  \nonumber \\
& + &  \frac{1}{\epsilon}\;
 \frac{1}{st} \biggl[
-3\; {\rm L}^2  +2 \;{\rm L}\;\ln(x)-\zeta_2
\biggr] \nonumber \\
  & + & \frac{1}{3st} \biggl\{
-10\;{\rm L}^3  + 12\;{\rm L}^2\;\ln(x) + 51\;{\rm L}\;\zeta_2\nonumber \\
&-&21\;\zeta_3 - 42\;\zeta_2\;\ln(x) - 2\;\ln^3(x) +
36\;\zeta_2\;\ln(1 + x) \nonumber \\
&+&  6\;\ln^2(x)\;\ln(1 + x) + 12\;\ln(x)\;{\litwo}( -x)
- 12\;{\litr}( -x)
\biggr\} \nonumber \\
 &+& \epsilon \; \frac{1}{3st} 
 \biggl\{
-5\;{\rm L}^4 + 10\;{\rm L}^3\;\ln(x) + 111\;{\rm L}^2\;\zeta_2  \nonumber \\
&+& {\rm L}\;(104\;\zeta_3 - 126\;\zeta_2\;\ln(x) - 6\;\ln^3(x)
+ 108\;\zeta_2\;\ln(1 + x) \nonumber \\
&+& 18\;\ln^2(x)\;\ln(1 + x) + 36\;\ln(x)\;{\litwo}( -x) - 36\;{\litr}( -x))
\nonumber \\
&-&372\;\zeta_4 - 78\;\zeta_3\;\ln(x) + 30\;\zeta_2\;\ln^2(x)
+ 2\;\ln^4(x) + 12\;\zeta_3\;\ln(1 + x)
\nonumber \\
&-& 60\;\zeta_2\;\ln(x)\;\ln(1 + x) - 8\;\ln^3(x)\;\ln(1 + x)
+ 18\;\zeta_2\;\ln^2(1 + x)
\nonumber \\
&+& 3\;\ln^2(x)\;\ln^2(1 + x) +
 12\;\ln(x)\; \liot 
 ( -x)  \nonumber \\
&-&6\;(4\;\zeta_2 + 3\;\ln^2(x) - 2\;\ln(x)\;\ln(1 + x))
\;{\litwo}( -x) \nonumber \\
&+& 24\;\ln(x)\;{\litr}( -x) - 12\;\ln(1 + x)\;{\litr}( -x)
\nonumber \\
&-& 12\;{\lifo}( -x) - 12\;{\litt}(-x)
\biggr\}
\label{B5l2m3d2ms}
\end{eqnarray}
\begin{eqnarray}
{\tt B5l2m3(k_2\cdot p_2)} &=&
 \frac{1}{4}\left(\frac{s}{u}\right)^2 \biggl\{
{\rm L}^2\;(6\;x\;\zeta_2 + 2\;x\;\ln(x) + 2\;x^2\;\ln(x)
+ x\;\ln^2(x)) \nonumber \\
&+& {\rm L}\;(16\;x\;\zeta_2 - 8\;x^2\;\zeta_2 - 4\;x\;\zeta_3 -
2\;\ln(x) + 2\;x^2\;\ln(x) \nonumber \\
&+& 4\;x\;\zeta_2\;\ln(x) + 2\;x\;\ln^2(x) - 2\;x^2\;\ln^2(x) -
   12\;x\;\zeta_2\;\ln(1 + x) \nonumber \\
&-& 2\;x\;\ln^2(x)\;\ln(1 + x)
- 4\;x\;\ln(x)\;{\litwo}( -x) + 4\;x\;{\litr}( -x)) \biggr\} \nonumber \\
 &+& \frac{1}{120} \left(\frac{s}{u}\right)^2 \biggl\{ +120\;\zeta_2 + 360\;x\;\zeta_2
- 120\;x^2\;\zeta_2 - 1560\;x\;\zeta_4 - 480\;x\;\zeta_3
\nonumber \\
&-& 240\;x^2\;\zeta_3 - 240\;x\;\zeta_2\;\ln(x)
- 480\;x^2\;\zeta_2\;\ln(x) - 360\;x\;\zeta_3\;\ln(x)
\nonumber \\
&+& 30\;\ln^2(x) + 60\;x\;\ln^2(x) - 30\;x^2\;\ln^2(x) -
 180\;x\;\zeta_2\;\ln^2(x) \nonumber \\
&-& 20\;x\;\ln^3(x) - 20\;x^2\;\ln^3(x) - 5\;x\;\ln^4(x)
+ 720\;x^2\;\zeta_2\;\ln(1 + x)
\nonumber \\
&+& 120\;x\;\zeta_3\;\ln(1 + x) - 120\;x\;\zeta_2\;\ln(x)\;\ln(1 + x) +
120\;x^2\;\ln^2(x)\;\ln(1 + x) \nonumber \\
&+&
 180\;x\;\zeta_2\;\ln^2(1 + x) + 30\;x\;\ln^2(x)\;\ln^2(1 + x) +
120\;x\;\ln(x)\;\liot 
(-x) \nonumber \\
&+& 60\;x\;(-8\;\zeta_2 - \ln^2(x) + 2\;\ln(x)\;
(2\;x + \ln(1 + x)))\;{\litwo}( -x)
\nonumber \\
&-& 240\;x^2\;{\litr}( -x) +
 240\;x\;\ln(x)\;{\litr}( -x) - 120\;x\;\ln(1 + x)\;{\litr}( -x)
\nonumber \\
&-& 360\;x\;{\lifo}( -x) - 120\;x\;{\litt}(-x)
\biggr\}
\label{B5l2m2N1ms}
\end{eqnarray}
\begin{eqnarray}
{\tt B5l2m3(k_2\cdot p_3)} &=&
 \frac{1}{4}\left(\frac{s}{u}\right)^2 \biggl\{
{\rm L}^2\;(12\;x\;\zeta_2 + 6\;x^2\;\zeta_2 - 2\;\ln(x) -
2\;x\;\ln(x) + 2\;x\;\ln^2(x)  \nonumber \\
&+& x^2\;\ln^2(x)) +  {\rm L}\;(-4\;\zeta_2 + 20\;x\;\zeta_2 -
8\;x\;\zeta_3 - 4\;x^2\;\zeta_3 \nonumber \\
&-&  2\;\ln(x) + 2\;x^2\;\ln(x) + 8\;x\;\zeta_2\;\ln(x) +
4\;x^2\;\zeta_2\;\ln(x) \nonumber \\
&+& 4\;x\;\ln^2(x) - 24\;x\;\zeta_2\;\ln(1 + x) - 12\;x^2\;\zeta_2\;
\ln(1 + x) \nonumber \\
&-& 4\;x\;\ln^2(x)\;\ln(1 + x) - 2\;x^2\;\ln^2(x)\;\ln(1 + x) \nonumber \\
&-& 4\;x\;(2 + x)\;\ln(x)\;{\litwo}( -x) + 4\;x\;(2 + x)\;{\litr}( -x))
\biggr\} \nonumber \\
&+&\frac{1}{24} \left(\frac{s}{u}\right)^2 \biggl\{
-24\;(-1 + (-3 + x)\;x)\;\zeta_2 + 72\;\zeta_3 + 24\;x\;\zeta_3 -
552\;x\;(2 + x)\;\zeta_4 \nonumber \\
&-& 48\;x\;(2 + x)\;\zeta_3\;\ln(x) + \ln(x)\;(48\;(1 + x)\;\zeta_2 \nonumber \\
&+& \ln(x)\;(6 - 6\;x\;(-2 + x + 16\;\zeta_2 + 8\;x\;\zeta_2)
+ 4\;(1 + x)\;\ln(x)
\nonumber \\
&-& x\;(2 + x)\;\ln^2(x))) + 24\;x\;(2 + x)\;\zeta_3\;\ln(1 + x) \nonumber \\
&+&
 4\;(-3\;(1 + 3\;x)\;(6\;\zeta_2 + \ln^2(x)) + \ln(-x)\;(24\;x\;(2 + x)\;
\zeta_2
\nonumber \\
&+& 3\;\ln(x)\;(2 + 6\;x + x\;(2 + x)\;\ln(x))))\;\ln(1 + x) \nonumber \\
&+& 6\;x\;(2 + x)\;(6\;\zeta_2 - 2\;\ln(-x)\;\ln(x) + \ln^2(x))\;\ln^2(1 + x)
\nonumber \\
&+&
 96\;x\;(2 + x)\;\zeta_2\;(\zeta_2 - \ln(-x)\;\ln(1 + x) - {\litwo}( -x))
\nonumber \\
&+& 24\;\ln(x)\;(\zeta_2 - \ln(-x)\;\ln(1 + x) - {\litwo}( -x))
\nonumber \\
&+& 72\;x\;\ln(x)\;(\zeta_2 - \ln(-x)\;\ln(1 + x) - {\litwo}( -x)) \nonumber \\
&+&
 12\;x\;(2 + x)\;\ln^2(x)\;(\zeta_2 - \ln(-x)\;\ln(1 + x) - {\litwo}( -x))
\nonumber \\
&+& 24\;{\litr}( -x) + 72\;x\;{\litr}( -x) +
48\;x\;(2 + x)\;\ln(x)\;{\litr}( -x) \nonumber \\
&-& 
 24\;x\;(2 + x)\;\ln(1 + x)\;{\litr}( -x) - 12\;x\;(2 + x)\;
(2\;\zeta_3\;\ln(x)
\nonumber \\
&+& 2\;\zeta_2\;\ln(x)\;\ln(1 + x) - \ln(-x)\;\ln(x)\;\ln^2(1 + x)
- 2\;\ln(x)\;\liot 
( -x) \nonumber \\
&-& 
2\;\ln(x)\;\ln(1 + x)\;{\litwo}( -x) + 6\;{\lifo}( -x) + 2\;{\litt}(-x))
\biggr\}
\label{B5l2m2N2ms}
\end{eqnarray}

\begin{eqnarray}
{\tt B5l3m} = & + &
 \frac{1}{6u} \biggl\{
{\rm L}^2\;(-18\;\zeta_2 - 3\;\ln^2(x)) + {\rm L}\;(12\;\zeta_3 +
2\;\ln^3(x) + 36\;\zeta_2\;\ln(1 + x)
\nonumber \\
&+& 6\;\ln^2(x)\;\ln(1 + x) + 12\;\ln(x)\;{\litwo}( -x) - 12\;{\litr}( -x))
\nonumber \\
&-& 81\;\zeta_4 - 12\;\zeta_3\;\ln(1 + x) - 2\;\ln^3(x)\;\ln(1 + x) -
18\;\zeta_2\;\ln^2(1 + x)
\nonumber \\
&-& 3\;\ln^2(x)\;\ln^2(1 + x) - 12\;\ln(x)\;\liot 
(-x) - 6\;\ln(x)\;(\ln(x)
\nonumber \\
&+& 2\;\ln(1 + x))\;{\litwo}( -x) +
 12\;\ln(x)\;{\litr}( -x) + 12\;\ln(1 + x)\;{\litr}( -x)
\nonumber \\
&-& 12\;{\lifo}( -x) + 12\;{\litt}(-x)
\biggr\}
\label{B5l3mms}
\\
{\tt B5l3md3} = & + & \frac{1}{\epsilon}\;
 \frac{1}{st} \biggl[
-{\rm L}^2 + {\rm L}\;\ln(x) \biggr] \nonumber \\
&+&\;
\frac{m_s^2}{3 x} \biggl\{ -8\;{\rm L}^3 + 12\;{\rm L}^2\;\ln(x) +
{\rm L}\;(24\;\zeta_2 - 3\;\ln^2(x))
\nonumber \\
&-&21\;\zeta_2\;\ln(x) - \ln^3(x) + 18\;\zeta_2\;\ln(1 + x) +
3\;\ln^2(x)\;\ln(1 + x)
\nonumber \\
&+& 6\;\ln(x)\;{\litwo}( -x) - 6\;{\litr}( -x)
\biggr\}
\end{eqnarray}

\begin{eqnarray}
{\tt B5l3m(k_1\cdot p_2)} = & + &
\frac{1}{12} \left(\frac{s}{u}\right)^2 \biggl\{ \nonumber \\
&+&{\rm L}^2\;(-18\;x\;\zeta_2 + 6\;\ln(x) + 6\;x\;\ln(x) - 3\;x\;\ln(x)^2)
\nonumber \\
&+& {\rm L}\;(36\;\zeta_2 - 36\;x\;\zeta_2 + 12\;x\;\zeta_3 + 6\;\ln(x) - 6\;x^2\;\ln(x)
\nonumber \\
&-& 12\;x\;\ln(x)^2 + 2\;x\;\ln(x)^3 + 36\;x\;\zeta_2\;\ln(1 + x) +
   6\;x\;\ln(x)^2\;\ln(1 + x) \nonumber \\
&+& 12\;x\;\ln(x)\;{\litwo}( -x) - 12\;x\;{\litr}( -x))
+12\;\zeta_2 - 12\;x\;\zeta_2
\nonumber \\
&+& 12\;x^2\;\zeta_2 - 12\;\zeta_3 + 12\;x\;\zeta_3 - 81\;x\;\zeta_4 - 3\;\ln(x)^2 - 6\;x\;\ln(x)^2
\nonumber \\
&+& 3\;x^2\;\ln(x)^2 - 2\;\ln(x)^3 + 2\;x\;\ln(x)^3 - 36\;\zeta_2\;\ln(1 + x)
\nonumber \\
&+& 36\;x\;\zeta_2\;\ln(1 + x) -
 12\;x\;\zeta_3\;\ln(1 + x) - 6\;\ln(x)^2\;\ln(1 + x) \nonumber \\
&+& 6\;x\;\ln(x)^2\;\ln(1 + x) - 2\;x\;\ln(x)^3\;\ln(1 + x)
- 18\;x\;\zeta_2\;\ln(1 + x)^2 \nonumber \\
&-& 3\;x\;\ln(x)^2\;\ln(1 + x)^2 - 12\;x\;\ln(x)\;\liot 
( -x) -
 6\;\ln(x)\;(2 - 2\;x + x\;\ln(x) \nonumber \\
&+& 2\;x\;\ln(1 + x))\;{\litwo}( -x)
+ 12\;{\litr}( -x) - 12\;x\;{\litr}( -x) + 12\;x\;\ln(x)\;{\litr}( -x)
\nonumber \\
&+& 12\;x\;\ln(1 + x)\;{\litr}( -x) - 12\;x\;{\lifo}( -x) +
12\;x\;{\litt}(-x)
\biggr\}
\label{B5l3mN1ms}
\end{eqnarray}

\begin{eqnarray}
{\tt B5l3m(k_1\cdot p_3)} = & + &
 \frac{1}{12} \left(\frac{s}{u}\right)^2  \biggl\{ \nonumber \\
&+& {\rm L}^2\;(18\;x^2\;\zeta_2 - 6\;x\;\ln(x) - 6\;x^2\;\ln(x) + 3\;x^2\;\ln(x)^2)
\nonumber \\
&+& {\rm L}\;(-72\;x\;\zeta_2 - 12\;x^2\;\zeta_3 + 6\;\ln(x) - 6\;x^2\;\ln(x) - 6\;x\;\ln(x)^2
\nonumber \\
&+& 6\;x^2\;\ln(x)^2 - 2\;x^2\;\ln(x)^3 -
   36\;x^2\;\zeta_2\;\ln(1 + x) - 6\;x^2\;\ln(x)^2\;\ln(1 + x)
\nonumber \\
&-& 12\;x^2\;\ln(x)\;{\litwo}( -x) + 12\;x^2\;{\litr}( -x))
-12\;\zeta_2 - 60\;x\;\zeta_2 - 12\;x^2\;\zeta_2
\nonumber \\
&+& 24\;x\;\zeta_3 + 81\;x^2\;\zeta_4 - 12\;x\;\zeta_2\;\ln(x) - 12\;x^2\;\zeta_2\;\ln(x) - 3\;\ln(x)^2
\nonumber \\
&-& 6\;x\;\ln(x)^2 + 3\;x^2\;\ln(x)^2 + 2\;x\;\ln(x)^3 - 2\;x^2\;\ln(x)^3
\nonumber \\
&+&
 72\;x\;\zeta_2\;\ln(1 + x) + 12\;x^2\;\zeta_3\;\ln(1 + x) + 12\;x\;\ln(x)^2\;\ln(1 + x)
\nonumber \\
&+& 2\;x^2\;\ln(x)^3\;\ln(1 + x) + 18\;x^2\;\zeta_2\;\ln(1 + x)^2
+ 3\;x^2\;\ln(x)^2\;\ln(1 + x)^2 \nonumber \\
&+&
 12\;x^2\;\ln(x)\;\liot 
 ( -x) + 6\;x\;\ln(x)\;(4 + x\;\ln(x) +
2\;x\;\ln(1 + x))\;{\litwo}( -x)
\nonumber \\
&-&  24\;x\;{\litr}( -x) - 12\;x^2\;\ln(x)\;{\litr}( -x)
- 12\;x^2\;\ln(1 + x)\;{\litr}( -x) \nonumber \\
&+&
 12\;x^2\;{\lifo}( -x) - 12\;x^2\;{\litt}(-x)
\biggr\}
\label{B5l3mN2ms}
\end{eqnarray}

\begin{eqnarray}
{\tt B5l3m(k_2\cdot p_2)} =
& + &
\frac{1}{12} \left(\frac{s}{u}\right)^2 \biggl\{ \nonumber \\
&+& {\rm L}^2\;(18\;x\;\zeta_2 + 6\;x\;\ln(x) + 6\;x^2\;\ln(x) + 3\;x\;\ln^2(x))
\nonumber \\
&+& {\rm L}\;(36\;x\;\zeta_2 - 36\;x^2\;\zeta_2 - 12\;x\;\zeta_3 - 6\;\ln(x) +
6\;x^2\;\ln(x) - 12\;x^2\;\ln^2(x) \nonumber \\
&-& 2\;x\;\ln^3(x) - 36\;x\;\zeta_2\;\ln(1 + x)
- 6\;x\;\ln^2(x)\;\ln(1 + x) \nonumber \\
&-& 12\;x\;\ln(x)\;{\litwo}(-x) +
12\;x\;{\litr}(-x))-12\;\zeta_2 + 12\;x\;\zeta_2 \nonumber \\
&-& 12\;x^2\;\zeta_2 - 12\;x\;\zeta_3 + 12\;x^2\;\zeta_3 + 81\;x\;\zeta_4 + 3\;\ln^2(x)
\nonumber \\
&+& 6\;x\;\ln^2(x) - 3\;x^2\;\ln^2(x)
- 2\;x\;\ln^3(x) + 2\;x^2\;\ln^3(x) - 36\;x\;\zeta_2\;\ln(1 + x)
\nonumber \\
&+&
 36\;x^2\;\zeta_2\;\ln(1 + x) + 12\;x\;\zeta_3\;\ln(1 + x) - 6\;x\;\ln^2(x)\;\ln(1 + x) \nonumber \\
&+& 6\;x^2\;\ln^2(x)\;\ln(1 + x) + 2\;x\;\ln^3(x)\;\ln(1 + x) +
18\;x\;\zeta_2\;\ln^2(1 + x) \nonumber \\
&+& 3\;x\;\ln^2(x)\;\ln^2(1 + x) +
 12\;x\;\ln(x)\;\liot 
 ( -x) +6\;x\;\ln(x)\;(\ln(x)
\nonumber \\
&+& 2\;(-1 + x + \ln(1 + x)))\;
{\litwo}(-x) + 12\;x\;{\litr}(-x) - 12\;x^2\;{\litr}(-x) \nonumber \\
&-& 12\;x\;\ln(x)\;{\litr}(-x) -
 12\;x\;\ln(1 + x)\;{\litr}(-x) \nonumber \\
&+& 12\;x\;{\lifo}( -x) - 12\;x\;
{\litt}( -x)
\biggr\}
\label{B5l3mN3ms}
\end{eqnarray}

\begin{eqnarray}
{\tt B5l4m}  =
&+& \frac{1}{\epsilon^2} \frac{1}{s}\;[{\rm L}]
\nonumber \\
&-& \frac{1}{\epsilon} \; \frac{1}{s} \;\biggl[
-3/2 \;{\rm L}^2 + \;{\rm L}\;(-2 + \ln(x))+\;\zeta_2 \biggr]
 \nonumber \\
&-& \frac{1}{6s} \biggl\{
-7\;{\rm L}^3 + {\rm L}^2\;(-18 + 9\;\ln(x)) + {\rm L}\;(-24 + 24\;\zeta_2
+ 12\;\ln(x) - 3\;\ln^2(x))
\nonumber \\
&+&
12\;\zeta_2 + 12\;\zeta_3 - 24\;\zeta_2\;\ln(x) - \ln^3(x) +
18\;\zeta_2\;\ln(1 + x)
\nonumber \\
&+&  3\;\ln^2(x)\;\ln(1 + x) + 6\;\ln(x)\;{\litwo}( -x) - 6\;{\litr}( -x)
\biggr\}
\end{eqnarray}

\begin{eqnarray}
{\tt B5l4md}
 &=& \frac{1}{6 st} \biggl\{
-11\;{\rm L}^3 + 21\;{\rm L}^2\;\ln(x) + {\rm L}\;(30\;\zeta_2 - 9\;\ln^2(x)) \nonumber \\
&-&24\;\zeta_2\;\ln(x) - \ln^3(x) + 18\;\zeta_2\;\ln(1 + x) +
3\;\ln^2(x)\;\ln(1 + x)
\nonumber \\
&+& 6\;\ln(x)\;{\litwo}( -x) - 6\;{\litr}( -x)
\biggr\}
\end{eqnarray}

\section{Conclusions}

With the help of a direct expansion of MB integrals, we obtained all of
the planar box masters to leading order in the expansion in
the small fermion mass. The calculation has been substantially
simplified by the fact that all nontrivial integrations were
one-dimensional. Unfortunately, the remaining, non-planar MIs
introduce two further complications. First, the final integrals are
not directly one-dimensional. This is {\it per se} not an
insurmountable obstacle, since a similar situation has been dealt with
during the evaluation of the fully massless masters. However, in our
case it seems non-trivial to bring the integrals to the form of
harmonic sums. The second problem is connected to the fact, that
MB representations for the non-planar case involve all three kinematic
invariants. All in all, despite the above problems we hope that the
technique used in this paper will prove successful also in this case,
and we will be able to provide the complete Bhabha scattering cross
section in the limit of high energy and fixed angle from a direct
diagrammatic calculation and thus verify results obtained by matching
between the massive and massless cases \cite{Penin}. Moreover, our
approach can be applied to problems with more scales, in particular,
we have applied it to the integrals occuring in the Bhabha cross
section with two fermion species \cite{new}.

\section{Acknowledgments}

We would like to thank Sven Moch for useful discussions.
Work supported in part by Sonderforschungsbereich/Transregio 9--03 of DFG
`Computergest{\"u}tzte Theo\-re\-ti\-sche Teil\-chen\-phy\-sik',  by
the Sofja Kovalevskaja Award of the Alexander von Humboldt Foundation
sponsored by the German Federal Ministry of Education and Research,
and by the Polish State Committee for Scientific Research (KBN),
research projects in 2004--2005.

\appendix

\section{Mellin-Barnes representations for 5-line integrals}

This appendix contains the MB representations that we directly derived
for the 5-line integrals with as well without irreducible
numerators. Notice that the momentum $p_e$ can be any of the external
momenta, because we performed the integration with a single loop
momentum in the numerator with vector index kept free. As far as the
notation is concerned, the $a_i$ denote the power of line $i$. The
numbering can be read off Fig.~\ref{5lin}.

\begin{eqnarray}
{\tt B5l2m2} &=&
   \frac{ (-1)^{a_{12345}} e^{2\ep\gamma_E}}{
\prod_{j=1}^5\Gamma[a_{i}]
   \Gamma[4  - 2\ep- a_{13} ](2\pi i)^3}
 \int_{-i \infty}^{+i \infty} d \alpha \int_{-i \infty}^{+i \infty} d \beta \int_{-i \infty}^{+i \infty} d \gamma
\\\nonumber&&
   (-s)^{2  - \ep - a_{245}- \gamma - \alpha + \beta}
   (-t)^{\alpha}
\Gamma[-2  + \ep+ a_{13} + \beta]
   \\\nonumber&&\Gamma[-\gamma]
\Gamma[2  - \ep - a_{245}- \gamma - \alpha]
   \Gamma[-\alpha]
  \\\nonumber&&
   \Gamma[a_{2} + \alpha]
  \Gamma[a_4 + \alpha]\Gamma[4 - 2\ep- a_{113}  - \beta]
 \\\nonumber &&
\Gamma[-2 + \ep + a_{245} + \gamma + \alpha - \beta]
   \Gamma[a_{1} + \beta]
   \\\nonumber&&
\frac{\Gamma[4 - 2\ep- a_{2245}  - 2\alpha + \beta]
     \Gamma[2- \ep  - a_{24} - \gamma - \alpha + \beta]}
{\Gamma[4 - 2\ep- a_{245}  + \beta]
   \Gamma[4 - 2\ep- a_{22445}  - 2\gamma - 2\alpha + \beta]}
\end{eqnarray}

\begin{eqnarray}
{\tt B5l2m2(p_e \cdot k_2)} &=&
\frac{ (-1)^{a_{12345}-1} e^{2\ep \gamma_E}}
{
\Gamma[a_1] \Gamma[a_2] \Gamma[a_3] \Gamma[a_4]\Gamma[a_5]
\Gamma[4 - 2 \epsilon - a_{13} ](2\pi i)^3
}
\int_{-i \infty}^{+i \infty} d \alpha \int_{-i \infty}^{+i \infty}
d \beta \int_{-i \infty}^{+i \infty} d \gamma
\nonumber \\
&&(-s)^{(2 - \epsilon - a_{245} - \alpha + \beta  - \gamma)}
\;(-t)^\gamma
\Gamma[-\alpha] \Gamma[-\gamma] \;\Gamma[a_1 + \beta]
\nonumber \\&&
\frac{\Gamma[4 - 2 \epsilon- a_{113} - \beta ]
  \Gamma[-2 + \epsilon + a_{13} + \beta ]}
{ \Gamma[5  - 2 \epsilon - a_{245} + \beta] }
\nonumber \\
&& \frac{\Gamma[-2  + \epsilon + a_{245} + \alpha - \beta + \gamma]}
{\Gamma[4  - 2 \epsilon -  a_{22445} - 2 \alpha + \beta - 2 \gamma]}
\frac{\Gamma[2  - \epsilon - a_{245} - \alpha - \gamma]}
{\Gamma[5 - 2 \epsilon - a_{22445} - 2 \alpha + \beta  - 2 \gamma]}
\nonumber \\
&&  \biggl[
p_e \cdot (p_1 +  p_2)
\Gamma[5 - 2 \epsilon - a_{22445} + \beta  - 2 \gamma]
\nonumber \\&&
\Gamma[4  - 2 \epsilon - a_{22445} - 2 \alpha + \beta- 2 \gamma]
\nonumber \\
&&
\Gamma[a_2 + \gamma]
\Gamma[a_4 + \gamma]
\Gamma[3 - \epsilon - a_{24} - \alpha + \beta - \gamma]
 \nonumber \\&&
+~
\Gamma[4- 2 \epsilon -  a_{224445} + \beta  - 2 \gamma]
\Gamma[5 - 2 \epsilon - a_{22445} - 2 \alpha + \beta  - 2 \gamma]
\nonumber \\ &&
\Gamma[2 - \epsilon - a_{24} - \alpha + \beta - \gamma]
\nonumber \\ &&
\biggl( p_e \cdot p_1
\;\Gamma[1 + a_2 + \gamma] \;\Gamma[a_4 + \gamma]
+ p_e \cdot p_3 \;\Gamma[a_2 + \gamma] \;\Gamma[1 + a_4 + \gamma] \biggr)
\biggr]
\label{numB5l2m2}
\end{eqnarray}

\begin{eqnarray}
{\tt B5l2m3(p_e \cdot k_2)} &=&
\frac{ (-1)^{a_{12345}}\;
e^{2\ep\gamma_E}}{ \prod_{j=1}^5\Gamma[a_{i}]
   \Gamma[4  - 2\ep- a_{123} ](2\pi i)^4}
\int_{-i \infty}^{+i \infty} d \alpha \int_{-i \infty}^{+i \infty}
d \beta \int_{-i \infty}^{+i \infty} d \gamma \int_{-i \infty}^{+i \infty} d \delta
\nonumber \\ &&
(-s)^{\gamma} \;(-t)^{(4 - 2 \epsilon - a_{12345} - \beta - \delta  - \gamma)} \nonumber
\\
&& \frac{\Gamma[-\beta] \;\Gamma[-\gamma] \;\Gamma[-\delta] \; \Gamma[a_3 + \alpha + 2 \;\beta] \; \Gamma[2 - \epsilon - a_{45} + \alpha - \delta  - \gamma]}
{\Gamma[7 - 3 \epsilon- a_{12345} - \beta]  } \nonumber \\
&&
\frac{\Gamma[2 - \epsilon - a_{13} - \beta ]
\;\Gamma[2- \epsilon - a_{23} - \alpha - \beta ]}
{ \Gamma[a_5 - \alpha + 2 \;\gamma] \;\Gamma[1 + a_5 - \alpha + 2 \;\gamma]} \nonumber \\
&&
  \Gamma[-4 + 2 \;\epsilon + a_{12345} + \beta + \delta + \gamma] \nonumber \\
&& \biggl\{ \Gamma[4 - 2 \;\epsilon - a_{1235} - \beta - \delta  - \gamma] \nonumber \\
&&
   \biggl[ (p_e \cdot p_2) \;\Gamma[1 + a_5 + \gamma] \;\Gamma[-\alpha + \gamma] - (p_e \cdot p_1) \;\Gamma[a_5 + \gamma] \;\Gamma[1 - \alpha + \gamma] \biggr]
\nonumber \\
&&\Gamma[a_5 - \alpha + 2 \;\gamma] \;\Gamma[1 + a_5 - \alpha + 2 \;\delta + 2 \;\gamma] \nonumber \\
&&+
   [(p_e \cdot p_3) - (p_e \cdot p_1)] \;\Gamma[5- 2 \epsilon  - a_{1235} - \beta - \delta - \gamma] \nonumber \\
&&\Gamma[a_5 + \gamma] \;\Gamma[-\alpha + \gamma] \;\Gamma[1 + a_5 - \alpha + 2 \;\gamma] \;
    \Gamma[a_5 - \alpha + 2 \;(\delta + \gamma)] \biggr\}
\label{numB5l2m3}
\end{eqnarray}

\begin{eqnarray}
{\tt B5l3m(p_e \cdot k_2)} &=&
\frac{
(-1)^{a_{12345}}e^{2\ep\gamma_E}}
{\prod_{j=1}^5\Gamma[a_{i}] \Gamma[4 - 2\ep- a_{123}](2\pi i)^4}
 \int_{-i \infty}^{+i \infty} d \alpha \int_{-i \infty}^{+i \infty}
d \beta \int_{-i \infty}^{+i \infty} d \gamma \int_{-i \infty}^{+i \infty} d \delta
\nonumber \\&&
 (-s)^{4 - 2\ep- a_{12345} - \alpha  - \beta - \delta}(-t)^{\delta}
\nonumber \\ &&
 \Gamma[2 - \ep- a_{13} - \alpha  - \gamma]
  \Gamma[a_1 + \gamma]
\nonumber \\ &&
  \frac{\Gamma[-4 + 2\ep + a_{12345} + \alpha + \beta + \delta]}{\Gamma[6 - 3\ep- a_{12345} - \alpha ]}
\nonumber \\ &&
\frac{ \Gamma[-\alpha] \Gamma[-\beta]}{\Gamma[7  - 3\ep- a_{12345} - \alpha]}
  \Gamma[4 - 2\ep- a_{12345} - \alpha  - \gamma - \beta - \delta]
 \nonumber \\ &&
 \frac{\Gamma[-\delta]}{\Gamma[4 - 2\ep- a_{1123} - 2\alpha  - \gamma]}
 \nonumber \\ &&
 \frac{\Gamma[2  - \ep- a_{12} - \alpha]}{ \Gamma[8- 4\ep - a_{112233445} - 2\alpha  - \gamma - 2\beta - 2\delta]}
\nonumber \\ &&
  \frac{ \Gamma[4  - 2\ep- a_{1123}  - \gamma]
  }{\Gamma[9 - 4\ep- a_{112233445} - 2\alpha  - \gamma - 2\beta - 2\delta]}
 \nonumber \\ &&
 (-((p_e\cdot p_1)~\Gamma[7 - 3\ep- a_{12345} - \alpha ]
     \Gamma[8 - 4\ep- a_{112233445} - 2\alpha  - \gamma - 2\delta]
 \nonumber \\ &&
    \Gamma[9  - 4\ep - a_{112233445}- 2\alpha- \gamma - 2\beta - 2\delta]
     \Gamma[4 - 2\ep- a_{1234} - \alpha  - \beta - \delta]
 \nonumber \\ &&
    \Gamma[a_4 + \delta]
     \Gamma[-2+ \ep  + a_{123} + \alpha + \gamma + \delta])
 \nonumber \\ &&
    + \Gamma[6 - 3\ep- a_{12345} - \alpha ]
\nonumber \\ &&
    ((p_e\cdot(p_1 + p_2))~\Gamma[9  - 4\ep- a_{112233445} - 2\alpha - \gamma - 2\delta]
\nonumber \\ &&
      \Gamma[8 - 4\ep- a_{112233445} - 2\alpha  - \gamma - 2\beta - 2\delta]
 \nonumber \\ &&
     \Gamma[5  - 2\ep- a_{1234} - \alpha - \beta - \delta]
      \Gamma[a_4 + \delta]
 \nonumber \\ &&
     \Gamma[-2+ \ep  + a_{123} + \alpha + \gamma + \delta]
\nonumber \\ &&
      +
     \Gamma[8 - 4\ep- a_{112233445} - 2\alpha  - \gamma - 2\delta]
\nonumber \\ &&
      \Gamma[9  - 4\ep - a_{112233445} - 2\alpha- \gamma - 2\beta - 2*\delta]
 \nonumber \\ &&
     \Gamma[4 - 2\ep - a_{1234} - \alpha - \beta - \delta]
 \nonumber \\ &&
     ((p_e\cdot p_3)~\Gamma[1 + a_4 + \delta]
      \Gamma[-2 + \ep+ a_{123} + \alpha  + \gamma + \delta]
 \nonumber \\ &&
     +
      (p_e\cdot  p_1)~ \Gamma[a_4 + \delta]
       \Gamma[-1 + \ep  + a_{123} + \alpha+ \gamma + \delta])))
\end{eqnarray}

\begin{eqnarray}
{\tt B5l3m(p_e \cdot k_1)} &=&
\frac{ (-1)^{a_{12345}}\;
e^{2\ep\gamma_E}}{ \prod_{j=1}^5\Gamma[a_{i}]
   \Gamma[5  - 2\ep- a_{123} ](2\pi i)^4}
   \int_{-i \infty}^{+i \infty} d \alpha \int_{-i \infty}^{+i \infty}
d \beta \int_{-i \infty}^{+i \infty} d \gamma \int_{-i \infty}^{+i \infty} d \delta \nonumber \\
&&
(-s)^{(4 - 2 \;\epsilon)- a_{12345}- \alpha - \beta - \delta } \;(-t)^\delta \nonumber \\
&& \frac{\Gamma[-4 + 2 \;\epsilon+ a_{12345} + \alpha + \beta + \delta ]}
{\Gamma[6 - 3 \;\epsilon - a_{12345} - \alpha]} \nonumber \\
&& \frac{\Gamma[-\alpha]\;\Gamma[-\beta]}
{\Gamma[7- 3 \;\epsilon - a_{12345} - \alpha ]\;\Gamma[5 - 2 \;\epsilon- a_{123}]} \nonumber \\
&&\frac{\Gamma[-\delta]}{\Gamma[4- 2 \;\epsilon - a_{1123}  - 2 \;\alpha  - \gamma] \;\Gamma[5 - 2 \;\epsilon- a_{1123} - 2 \;\alpha - \gamma]} \nonumber \\
&& \frac{\Gamma[2 - \epsilon- a_{13} - \alpha  - \gamma]}
{ \Gamma[8 - 4 \;\epsilon- a_{112233445} - 2 \;\alpha - 2 \;\beta - 2 \;\delta  - \gamma]} \nonumber \\
&& \frac{\Gamma[4 - 2 \;\epsilon- a_{12345} - \alpha - \beta - \delta  - \gamma]}
{\Gamma[9 - 4 \;\epsilon- a_{112233445} - 2 \;\alpha - 2 \;\beta - 2 \;\delta  - \gamma]} \nonumber \\
&&  \biggl\{ (p_e \cdot p_3)  \;\Gamma[1 + a_4 + \delta] \; \Gamma[6- 3 \;\epsilon - a_{12345} - \alpha ]  \nonumber \\
&&
\Gamma[4 - 2 \;\epsilon- a_{1234} - \alpha - \beta - \delta ]
\;\Gamma[3 - \epsilon- a_{12} - \alpha ] \; \nonumber \\
&&    \Gamma[8 - 4 \;\epsilon- a_{112233445} - 2 \;\alpha - 2 \;\delta  - \gamma] \nonumber \\
&&
\Gamma[9 - 4 \;\epsilon- a_{112233445} - 2 \;\alpha - 2 \;\beta - 2 \;\delta  - \gamma] \nonumber \\
&&    \Gamma[5 - 2 \;\epsilon- a_{1123} - \gamma] \;\Gamma[4  - 2 \;\epsilon- a_{1123} - 2 \;\alpha - \gamma] \nonumber \\
&& \;\Gamma[a_1 + \gamma] \;\Gamma[-2 + \epsilon + a_{123} + \alpha + \delta + \gamma] +
   \Gamma[a_4 + \delta] \nonumber \\
&& \biggl[ -(p_e \cdot p_1)  \;\Gamma[7- 3 \;\epsilon - a_{12345} - \alpha ] \nonumber \\
&&
\Gamma[4 - 2 \;\epsilon- a_{1234} - \alpha - \beta - \delta ]
\nonumber \\
&&       \Gamma[8 - 4 \;\epsilon - a_{112233445} - 2 \;\alpha - 2 \;\delta - \gamma] \nonumber \\
&&
\Gamma[9- 4 \;\epsilon - a_{112233445} - 2 \;\alpha - 2 \;\beta - 2 \;\delta  - \gamma] \; \nonumber \\
&&       \biggl[ \Gamma[3 - \epsilon- a_{12} - \alpha ] \;\Gamma[5 - 2 \;\epsilon- a_{1123}  - \gamma] \nonumber \\
&& \Gamma[4 - 2 \;\epsilon- a_{1123} - 2 \;\alpha  - \gamma] \;\Gamma[a_1 + \gamma] \nonumber \\
&&+
        \Gamma[2 - \epsilon - a_{12} - \alpha] \;\Gamma[4  - 2 \;\epsilon- a_{1123}  - \gamma] \nonumber \\
&&
\Gamma[5- 2 \;\epsilon - a_{1123}- 2 \;\alpha  - \gamma] \;\Gamma[1 + a_1 + \gamma] \biggr] \; \nonumber \\
&&       \Gamma[-2+ \epsilon + a_{123} + \alpha + \delta  + \gamma] \nonumber \\
&&
+ \Gamma[6 - 3 \;\epsilon- a_{12345} - \alpha ] \;\Gamma[3 - \epsilon- a_{12} - \alpha ] \; \nonumber \\
&&      \Gamma[5- 2 \;\epsilon - a_{1123}  - \gamma] \;\Gamma[4 - 2 \;\epsilon - a_{1123} - 2 \;\alpha - \gamma]
\nonumber \\
&&     \Gamma[a_1 + \gamma]
 \biggl[ ((p_e \cdot (p_1 + p_2) ) \;\Gamma[5 - 2 \;\epsilon- a_{1234} - \alpha - \beta - \delta ] \nonumber \\
&&
\Gamma[9 -4 \;\epsilon- a_{112233445} - 2 \;\alpha - 2 \;\delta  - \gamma] \nonumber \\
&&        \Gamma[8 - 4 \;\epsilon- a_{112233445} - 2 \;\alpha - 2 \;\beta - 2 \;\delta  - \gamma] \nonumber \\
&& \Gamma[-2 + \epsilon+ a_{123} + \alpha + \delta  + \gamma] \nonumber \\
&& +
       (p_e \cdot p_1)  \;\Gamma[4 - 2 \;\epsilon- a_{1234} - \alpha - \beta - \delta ] \nonumber \\
&& \Gamma[8 - 4 \;\epsilon- a_{112233445} - 2 \;\alpha - 2 \;\delta  - \gamma] \nonumber \\
&&        \Gamma[9 - 4 \;\epsilon- a_{112233445} - 2 \;\alpha - 2 \;\beta - 2 \;\delta  - \gamma] \nonumber \\
&&\Gamma[-1 + \epsilon+ a_{123} + \alpha + \delta  + \gamma] \biggr] \biggr\}
\label{numB5l3m}
\end{eqnarray}

\end{document}